\documentclass{emulateapj} 
\usepackage{times}

\shorttitle{Millisecond Variability from XTE J1739-285}
\shortauthors{Kaaret et al.}

\begin{document}

\title{Evidence for 1122~Hz X-Ray Burst Oscillations from the
Neutron-Star X-Ray Transient XTE J1739-285}

\author{P.\ Kaaret\altaffilmark{1}, Z.\ Prieskorn\altaffilmark{1},
J.J.M.\ in 't Zand\altaffilmark{2}, S.\ Brandt\altaffilmark{3}, N.\
Lund\altaffilmark{3}, S.\ Mereghetti\altaffilmark{4}, D.\
G\"{o}tz\altaffilmark{5}, E.\ Kuulkers\altaffilmark{6}, J.A.\ Tomsick
\altaffilmark{7,8}}

\altaffiltext{1}{Department of Physics and Astronomy, University of
Iowa, Iowa City, IA, 52242; philip-kaaret@uiowa.edu,
zachary-prieskorn@uiowa.edu}

\altaffiltext{2}{SRON Netherlands Institute for Space Research,
Sorbonnelaan 2, 3584 CA Utrecht, the Netherlands;
J.J.M.in.t.Zand@sron.nl}

\altaffiltext{3}{Danish National Space Center, Juliane Maries Vej 30,
DK-2100 Copenhagen, Denmark; sb@dsri.dk, nl@dsri.dk}

\altaffiltext{4}{INAF - Istituto di Astrofisica Spaziale e Fisica
Cosmica, Milano, via Bassini 15, 20133 Milano, Italy;
sandro@iasf-milano.inaf.it}

\altaffiltext{5}{CEA - Saclay, DSM/DAPNIA/Service d'Astrophysique, Orme
des Merisiers, Bat.\ 709, F-91191 Gif-sur-Yvette, France;
diego@mi.iasf.cnr.it}

\altaffiltext{6}{SOC, ESA/ESAC, Urb.\ Villafranca del Castillo, PO Box
50727, 28080 Madrid, Spain; ekuulker@rssd.esa.int}

\altaffiltext{7}{Space Sciences Laboratory, 7 Gauss Way, University of
California, Berkeley, CA 94720; jtomsick@ssl.berkeley.edu}

\altaffiltext{8}{Center for Astrophysics and Space Sciences, Code 0424,
9500 Gilman Drive, University of California at San Diego, La Jolla, CA
92093}

\begin{abstract}

We report on millisecond variability from the X-ray transient XTE
J1739-285.  We detected six X-ray type I bursts and found evidence for
oscillations at $1122\pm 0.3$~Hz in the brightest X-ray burst.  Taking
into consideration the power in the oscillations and the number of
trials in the search, the detection is significant at the 99.96\%
confidence level.  If the oscillations are confirmed, the oscillation
frequency would suggest that XTE J1739-285 contains the fastest
rotating neutron star yet found.  We also found millisecond
quasiperiodic oscillations in the persistent emission with frequencies
ranging from 757~Hz to 862~Hz.  Using the brightest burst, we derive an
upper limit on the source distance of about 10.6~kpc.

\end{abstract}


\section{Introduction}

Weakly magnetized neutron stars can be spun up to rates of several
100~Hz by accretion in low-mass X-ray binaries \citep{Alpar82}.  The
first direct measurements of millisecond spin rates in actively
accreting neutron stars in low-mass X-ray binaries (LMXBs) came from
the discovery of oscillations in thermonuclear X-ray bursts occurring
on the neutron star surface \citep{Strohmayer96}.  The burst
oscillation frequencies were later found to be nearly equal to those of
coherent pulsations
\citep{intZand01,Strohmayer02,Chakrabarty03,Strohmayer03} implying that
the burst oscillation frequency indicates the neutron star spin rate. 
This X-ray technique has no known biases against the detection of very
high spin rates, unlike radio pulsation searches, and the sample of
X-ray burst oscillation frequencies has been exploited to constrain the
neutron star spin rate distribution.  Analysis of a sample of 11 X-ray
measured spin frequencies in the range 270-619~Hz suggests a limiting
spin rate near 760~Hz if the distribution is uniform and bounded
\citep{Chakrabarty03}.  This is below the expected maximum spin
frequency possible without centrifugal breakup and has been interpreted
as evidence that some physical process, possibly gravitational
radiation, limits the maximum possible spin rate.  However, the
discovery of a radio pulsar spinning at 716~Hz, a frequency above any
previously measured in X-rays, suggests that the true maximum spin rate
is higher \citep{Hessels06}.

Here, we describe observations made with the Rossi X-Ray Timing
Explorer (RXTE; Bradt, Rothschild, \& Swank 1993) following the
detection of X-ray bursts from the transient source XTE J1739-285 as
part of a program to search for millisecond oscillations in both X-ray
bursts and persistent emission from neutron star X-ray binaries which
are newly discovered or found to be active
\citep{Kaaret02,Kaaret03,Kaaret06}.  We detected six X-ray bursts and
found oscillations at a frequency of 1122~Hz in the brightest burst. 
This suggests that XTE J1739-285 contains the most rapidly rotating
neutron star yet discovered.  We describe our observations in \S 2, the
X-ray bursts in \S 3, and the persistent emission and the discovery of
kHz QPOs from the source in \S 4.  We discuss the results in \S 5.

\section{Observations of XTE J1739-285}

XTE J1739-285 is a transient neutron-star low-mass X-ray binary
(NS-LMXB) that was discovered during RXTE PCA scans of the Galactic
bulge on 1999 October 19 \citep{Markwardt99} and underwent short
outbursts in May 2001 and October 2003.  The source became active again
in August 2005 \citep{Bodaghee05} and two X-ray bursts were detected
with the JEM-X instrument on INTEGRAL on 2005 September 30 and  October
4 \citep{Brandt05}.  

Triggered by the detection of the X-ray bursts, we obtained 19
observations using RXTE in the period beginning 2005 October 12 and
ending 2005 November 16.  Data were obtained with the Proportional
Counter Array (PCA) in a spectral mode (Standard 2) with 256 energy
channels and  16~s time resolution, a low-resolution timing mode
(Standard 1) with no energy information and 0.125~s time resolution,
and a high-resolution timing mode (Event mode) with 122~$\mu$s time
resolution and 64 energy channels \citep{Jahoda06}.

\section{X-Ray Bursts}

We searched the Standard 1 data for X-ray bursts and found six, see
Table~\ref{bursttable}.  We examined their evolution by extracting
spectra for 0.25~s intervals of event mode data using all Proportional
Counter Units (PCUs) that were on during each burst and all layers. 
The fluxes were corrected for dead time effects with a maximum
correction of 5.5\%.  To remove the contribution of the persistent
emission, we subtracted off a spectrum from 10~s of data preceding each
burst.  Spectra were fitted in the 3-18~keV band (channels 3-24 in the
event mode data) with an absorbed blackbody model with the column
density fixed to $N_H = 7.5 \times 10^{21} \rm \, cm^{-2}$ which is the
hydrogen column density along the line of sight in the Milky Way
\citep{Dickey90}.  We found a burst-like event at Nov 11 10:27:22 UTC
which appears in only one of four PCUs on at the time and is likely a
detector breakdown event and not an X-ray burst.

\begin{deluxetable}{cccc}
\tablecolumns{3}
\tabletypesize{\scriptsize}
\tablecaption{Properties of X-Ray Bursts
  \label{bursttable}}
\tablewidth{0pt}
\tablehead{\colhead{\#}  & \colhead{Time} &  
           \colhead{Peak flux} &  \colhead{Decay}}
\startdata
1 & Oct 31 07:59:09 & $1.6 \pm 0.3$ & 5.0 \\
2 & Nov 4  11:34:23 & $2.8 \pm 0.5$ & 5.4 \\
3 & Nov 7  05:34:04 & $2.2 \pm 0.4$ & 5.0 \\
4 & Nov 7  07:30:53 & $1.1 \pm 0.3$ & 3.0 \\
5 & Nov 8  08:19:08 & $1.0 \pm 0.2$ & 3.1 \\
6 & Nov 11 09:47:10 & $1.2 \pm 0.3$ & 5.7 \\
\enddata

\tablecomments{The table lists for each burst: Time -- the time (UTC)
at the start of each burst (all bursts occurred in 2005); Peak flux --
the bolometric peak flux corrected for absorption in units of $10^{-8}
\rm \, erg \, cm^{-2} \, s^{-1}$; and Decay -- the decay ($e$-folding)
time in seconds.} \vspace{0.2in}
\end{deluxetable}

The brightest burst is \# 2, which had a peak bolometric flux, as
measured in 0.25~s intervals, of $(2.8 \pm 0.5) \times 10^{-8} \rm \,
erg \, cm^{-2} \, s^{-1}$.  The blackbody temperature rose rapidly
during the flux rise, reached a maximum near 2.5~keV, and then decayed
as the flux decayed.  This behavior is indicative of heating during the
rise and cooling during the decay and is characteristic of type I X-ray
bursts.  \citet{Kuulkers03} found that the empirical maximum bolometric
peak luminosities for photospheric radius expansion bursts from a
sample of 12 bursters located in globular clusters with known distances
was $3.8 \times  10^{38} \rm \, erg \, s^{-1}$, with an accuracy of
$\sim$15\%.  None of our bursts showed evidence of photospheric radius
expansion and, thus, their luminosities should be below this value. 
The peak flux of the brightest burst implies an upper limit on the
distance to XTE J1739-285 of about 10.6~kpc.  This is consistent with,
but somewhat closer than, the upper limit of 12~kpc from
\citet{Torres06}.

Bursts 1 and 4-6 have similar peak fluxes, while bursts 2 and 3 were
brighter, see Table~\ref{bursttable}.  Burst 3 had a rise time near
3~s, while the other rise times were near 1~s.  Burst 6 had a
relatively broad maximum extending over 5~s.  The decay ($e$-folding)
times are in the range 3--6~s.  The fact that the bursts have fast
rises and durations of tens of seconds suggests that they are typical
helium bursts.

\begin{figure}  
\centerline{\includegraphics[width=3.2in]{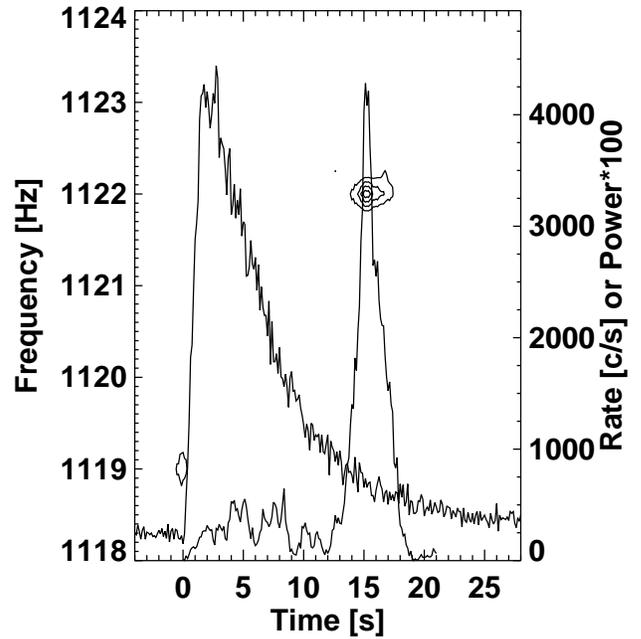}}
\caption{Dynamical power spectrum of burst 2 with the burst light curve
(count rate) and the power at 1122.0~Hz superposed. The contours are at
Leahy powers of 12, 20, 28, and 36.  The contours are generated from
power spectra for overlapping 4~s intervals of data with the power
plotted at the midpoint of the interval.  The curve which peaks near
2~s is the rate in the 3-18~keV band from all the PCUs on during the
observation (PCU0 and PCU2).  The curve which peaks near 15~s is the
Leahy power at 1122.0~Hz.  The scales for both curves are on the right
side.  The oscillation is significant at the 99.96\% confidence level.}
\label{burst2dyn} \end{figure}

Using high time resolution data, we computed power spectra for
overlapping 4~s intervals with 0.125~s between the starts of successive
intervals using events in PHA channels 4-24 for PCU0 and channels 3-24
all other PCUs.  The background, particularly at low energies, is
higher in PCU0 due to the loss of its propane layer.  We searched over
the duration of each burst defined as the interval where the total
counting rate exceeds the background plus persistent counting rate,
measured a few seconds before the burst, by at least a factor of 2.  We
searched for excess power in the range 20-2000~Hz, corresponding to
7920 frequency bins in the 4~s FFTs.  We found oscillations in the
second burst with a Leahy normalized power \citep{Leahy83} of 42.82 at
a frequency of $1122 \pm 0.3$~Hz, occurring in the burst decay 15~s
after the burst rise, see Fig.~\ref{burst2dyn}.  The four consecutive
FFTs around the one with the peak power all have powers above 40.31 at
1122.0~Hz.

To evaluate the significance of the signal, we used a Monte Carlo
simulation which generates Poisson distributed events following the
light curve from burst 2 in 0.125~s bins after smoothing with a moving
average over 9 bins.  The deadtime of the PCA is modeled by removing
any event which occurs within 10~$\mu$s after a previous event
\citep{Jahoda06}.  The number of events generated in each time bin is
larger than the observed counts so that after the deadtime correction
the number matches that in the actual light curve within Poisson
fluctuations.  We generated 400,000 trial bursts.  Each simulated burst
was analyzed using exactly the same analysis done on the real data,
specifically by calculating 4~s FFTs at overlapping intervals with
starts each 0.125 seconds and searching for power in the 20-2000~Hz
interval.  The chance probability of occurrence of the observed signal
is calculated by counting the fraction of trial bursts with powers
equal to or exceeding those observed.  This procedure eliminates the
ambiguity in estimating the equivalent number of independent trials for
the overlapping FFTs.  We found 58 bursts with at least one power above
42.82 of which 14 had at least 4 consecutive powers above 40.31 at the
same frequency.  We estimate the chance probability of occurrence of
the observed signal to be $3.5\times 10^{-5}$.  Allowing two trials for
two energy bands, the probability is $7.0\times 10^{-5}$ equivalent to
a $3.97\sigma$ confidence level.

The signal at 1122~Hz was detected in the brightest burst.  Since the
bursts are not a uniform sample and the presence of oscillations may
depend on burst properties and/or accretion history (e.g.\ Watts,
Strohmayer, Markwardt 2005), one can reasonably argue that the
significance should be evaluated for each burst individually,
particularly in this case since the brightest burst is the most likely
to give a detectable signal.  However, a more conservative approach is
to consider the full set of six bursts.  The durations of the bursts,
as defined above, are 21, 21, 22, 10, 11, and 18~s for bursts 1--6,
respectively.  To account for all the trials in all the bursts, we
multiply the number of trials by an additional factor of 103/21 = 4.9
for a chance probability of $3.4 \times 10^{-4}$.  The oscillation is
detected at a confidence level of 99.966\%, equivalent to a $3.6\sigma$
significance.

\begin{figure}
\centerline{\includegraphics[width=3.0in,angle=0]{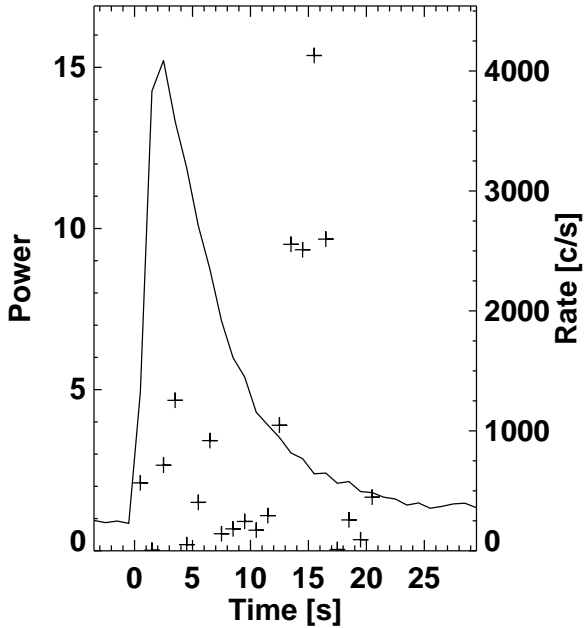}}
\caption{Power at 1122~Hz in independent 1~s intervals (crosses).  The
solid line is the count rate in the same 1~s intervals.  The powers at
1122~Hz used in the significance calculation are, in time order
starting at 12.5~s, 3.90, 9.51, 9.34, 15.37, and 9.67.  We note that
the power at 17.5~s at 1123~Hz is not shown on the plot and is 7.94.}
\label{burst2i} \end{figure}

To obtain an estimate of the significance of the oscillations which
does not depend on a simulation, we calculated FFTs in independent 4~s,
2~s, and 1~s intervals of data from burst 2.  There are five successive
1~s FFTs with power at 1122 Hz, see Fig.~\ref{burst2i}. The single
trial probability for having five powers at or above the observed
levels in the 1~s FFTs is $4.2 \times 10^{-11}$.  Accounting for all
sequential combinations of FFTs of fixed length of 1, 2, or 4~s with a
total duration of 5~s or less covering the 20-2000~Hz range in two
energy bands, we estimate a total of 605880 trials:
(17+18+19+20+21)$\times$1980$\times$2 + (9+10)$\times$3960$\times$2 +
5$\times$7920$\times$2.  The chance probability taking into account the
number of trials in the one burst is then $2.5 \times 10^{-5}$
($4.2\sigma$).  Accounting for the other bursts, the chance probability
is $1.2\times 10^{-4}$ ($3.8\sigma$).  This probability agrees within a
factor of 3 with that in the previous paragraph.

In the 4~s interval with the maximum power, the fractional rms
amplitude of the 1122.0~Hz oscillation is $0.13 \pm 0.02$.  The quality
factor, $Q = \nu/\Delta\nu$, of the oscillation in burst 2 is $Q >
1000$.  This is consistent with the values found in burst oscillations
and inconsistent with the values found in high frequency QPOs in the
persistent emission.  We searched for oscillations near half the
frequency.  The strongest signal has a Leahy power of 13.6 at 561.25~Hz
and occurs 7.75~s after the burst rise.  This is not a significant
signal; the chance probability of occurrence within the narrow window
searched, 562.5-558.5~Hz, is 0.09.  At the time of the maximum strength
of the 1122~Hz signal, the Leahy power near 561~Hz is less than 6.

\section{Persistent emission}

To study the persistent emission spectra, we used Standard-2 data from
PCU 2, removing data around the X-ray bursts, and estimating the
background using bright source background files.  The source was in a
relatively hard state, with a hard X-ray color, $HC$, defined as the
count rate in the 9.7--16~keV band divided by the 6.0--9.7~keV rate, of
$HC > 0.35$ for all observations before October 25.  The flux was
roughly constant and near $4 \times 10^{-10} \rm \, erg \, cm^{-2} \,
s^{-1}$ in the 2-20~keV band.  The spectra were adequately described by
the absorbed ($N_H = 7.5 \times 10^{21} \rm \, cm^{-2}$) sum of a
power-law with a photon index near 2 and a gaussian emission line with
a centroid of 6.7~keV and a flux consistent with that from Galactic
ridge emission at the source position \citep{Revnivtsev06}.  After
October 25, the source stayed in a softer state, $HC < 0.35$.  The flux
varied by a factor of two, with the maximum near $1.4 \times 10^{-9}
\rm \, erg \, cm^{-2} \, s^{-1}$ in the 2-20~keV band.  The spectra
showed distinct curvature at high energies and were adequately fitted
with the absorbed sum of a Comptonization model (compst in xspec) with
a temperature in the range 4--12~keV and an optical depth in the range
3--10, a blackbody with a temperature near 1.6~keV, and a gaussian
emission line with a centroid of 6.7~keV, a width less than 0.9~keV,
and an equivalent width of 100-300~eV, in some cases larger than the
expected Galactic ridge emission.  The blackbody component was not
required for some lower flux observations.  The spectra are similar to
those of other moderate luminosity NS-LMXBs \citep{Kaaret02,Kaaret03}.

\begin{deluxetable}{lcccc}
\tablecolumns{5}
\tabletypesize{\scriptsize}
\tablecaption{High frequency peaks in the persistent emission
  \label{qpotable}}
\tablewidth{0pt}
\tablehead{
  \colhead{Time}  & \colhead{Centroid} & 
     \colhead{Width} & \colhead{Amplitude} \\ 
  \colhead{(UTC)} & \colhead{(Hz)}     &
     \colhead{(Hz)}  & \colhead{(\%)} \\
    }
\startdata
Oct 31 06:22:11 & 837.9$\pm$0.3 & 5.3$\pm$0.9 & 1.7$\pm$0.2 \\
Nov 01 02:48:14 & 784.6$\pm$1.1 &  14$\pm$3   & 2.4$\pm$0.4 \\
Nov 08 04:37:47 & 814.4$\pm$3.2 &  36$\pm$9   & 3.1$\pm$0.6 \\
Nov 08 09:46:31 & 756.7$\pm$0.7 &   9$\pm$3   & 2.1$\pm$0.4 \\
Nov 10 08:30:35 & 861.8$\pm$0.7 &  13$\pm$2   & 2.7$\pm$0.3 \\
Nov 10 10:04:43 & 846.0$\pm$2.9 &  33$\pm$8   & 2.7$\pm$0.5 \\
\enddata
\tablecomments{The table includes: Time -- the UTC time at the
beginning of the observation, all observations were in 2005; Centroid
and Width - of the fitted Lorentzian; Amplitude - RMS fractional
amplitude of the fitted Lorentzian.}   
\end{deluxetable}

To further investigate the source state, we produced low frequency
power spectra using event mode data from all PCUs on during each
observation.  The sum power spectrum for observations with $HC > 0.35$
shows very-low-frequency noise (VLFN) and so-called high-frequency
noise (HFN).  The total noise power in the 0.01-100~Hz band is 1.0\%
fractional rms.  We fit the power spectrum with a model consisting of a
power-law to describe the VLFN and an exponentially cutoff power-law
for the HFN \citep{vdk95}.  The best fit parameters are a power law
index for the VLFN of $1.81 \pm 0.13$ and a power law index of $0.8 \pm
0.1$ and a cutoff frequency in the range of 20-60~Hz for the HFN.  The
weakness of the signal at high frequencies prevents an accurate
determination of the HFN parameters, but a HFN noise component is
required, $\Delta \chi^2 = 49$.  This power spectrum is consistent with
those in the ``lower banana'' state.  There is no detectable HFN in the
power spectrum from the observations with the softer hard color, $HC <
0.35$.  The total noise power in the 0.01-100~Hz band is 0.2\%
fractional rms and the spectrum is adequately described by a single
power law with an index of $1.08 \pm 0.02$.  The shape of the power
spectrum is consistent with those seen in the ``upper banana'' state,
but the total noise power is lower than usual and spectrally the state
would be classified as the ``lower banana'' \citep{vdk95}.

Based on the timing and color information, we identify XTE J1739-285 as
an atoll source.  The source appears to have been in the ``banana''
state during all of the observations analyzed here, although its
evolution along the ``banana'' appears somewhat unusual.

\begin{figure}
\centerline{\includegraphics[width=2.5in,angle=0]{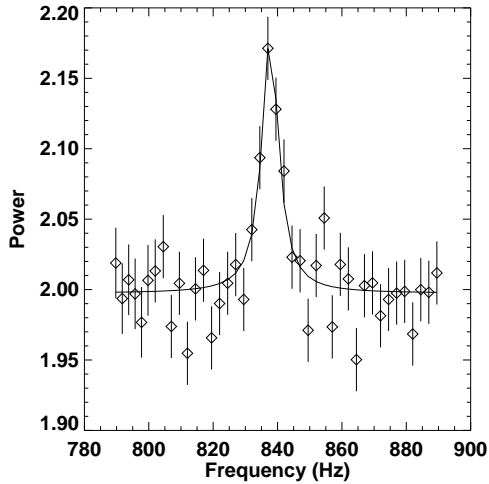}}
\caption{Power spectrum showing a kHz QPO from October 31.  The power
is Leahy normalized.} \label{khzqpo} \end{figure}

We searched for high frequency quasiperiodic oscillations (QPOs) in
each uninterrupted RXTE observation window.  We calculated averages of
2~s power spectra for PCA events in the 4.2-20.4~keV energy band
(channels 6-26 in the event mode data).  We searched for peaks in the
range from 100~Hz to 3000~Hz and fit any peak found with a Lorentzian
plus a constant equal to the calculated the Poisson noise level.  There
were several strong QPO detections, see Table~\ref{qpotable}.  The
second most significant detection occurred on October 31 and is shown
in Fig.~\ref{khzqpo}.  Allowing for 1102 trials, which is calculated by
dividing the search interval of 2900~Hz by the QPO trial widths of 5,
10, 20, 50, and 100~Hz, we estimate a chance probability of occurrence
of $1.2 \times 10^{-13}$ for this QPO.  The chance probabilities of
occurrence, taking into account the number of trials, of all the other
QPOs listed in the Table are less than $1 \times 10^{-4}$. These QPO
detections establish XTE J1739-285 as a new member of the class of
neutron-star low-mass X-ray binaries producing kHz QPOs.  We did not
detect two simultaneous kHz QPOs in any observation.  


\section{Discussion}

These observations establish that XTE J1739-285 exhibits millisecond
oscillations in its persistent X-ray emission.  The properties of the
source are generally similar to those of other atoll neutron-star X-ray
binaries, although the evolution of the timing noise in the banana
state is unusual.

The signal at 1122~Hz present in burst 2 is significant at the 99.96\%
confidence level.  If this signal represents a true burst oscillation
then it would be of substantial interest.  The near equality of the
burst oscillation frequency with the frequency of coherent pulsations
in the millisecond pulsars SAX J1808.4-3658
\citep{intZand01,Chakrabarty03} and XTE J1814-338 \citep{Strohmayer03}
and the frequency of coherent pulsations in a superburst from 4U
1636-536 \citep{Strohmayer02} strongly suggests that the burst
oscillation frequency indicates the neutron star spin frequency.  The
lack of any significant signal near 561~Hz in the burst from XTE
J1739-285 supports this interpretation and suggests that the possible
1122~Hz oscillation would be most naturally interpreted as the spin
rate of the neutron star.

If the burst oscillation frequency of 1122~Hz is the spin rate of the
neutron star, then XTE J1739-285 contains the most rapidly rotating
neutron star yet discovered.  This spin rate is close to the
centrifugal breakup limit for some equations of state of nuclear matter
\citep{Burgio03} and, therefore, may remove the motivation for a
physical limit on neutron star spin other than the centrifugal breakup
limit.  Furthermore, such a high spin rate would place constraints on
the nuclear equation of state, particularly if combined with a
measurement of the mass and/or radius of the neutron star.

\acknowledgments  

We thank Anna Watts and an anonymous referee for useful discussions and
the RXTE team, particularly Jean Swank and Evan Smith, for scheduling
these observations.  PK and ZP acknowledge support from NASA grant
NNGO5GM77G.




\begin{thebibliography}{}

\bibitem[Alpar et al.(1982)]{Alpar82} Alpar, M.A., Cheng, A.F.,
Ruderman, M. A., \& Shaham, J.\ 1982, Nature, 300, 728

\bibitem[Bodaghee et al.(2005)]{Bodaghee05} Bodaghee, A., Mowlavi, N.,
Kuulkers, E., Wijnands, R., et al. 2005, ATEL 592

\bibitem[Bradt, Rothschild, \& Swank(1993)]{Bradt93} Bradt, H.V.,
Rothschild, R.E., \& Swank, J.H. 1993, A\&AS, 97, 355

\bibitem[Brandt et al.(2005)]{Brandt05} Brandt, S., Kuulkers, E.,
Bazzano, A., Courvoisier, T. J.-L., Domingo, A., et al. 2005, ATEL 622

\bibitem[Burgio, Schulze, \& Weber(2003)]{Burgio03} Burgio, G.F,
Schulze, H.-J., Weber, F.\ 2003, A\&A, 408, 675

\bibitem[Chakrabarty et al.(2003)]{Chakrabarty03} Chakrabarty, D.,
Morgan, E.H., Muno, M.P., Galloway, D.K., Wijnands, R., van der Klis,
M., Markwardt, C.B.\ 2003, Nature, 424, 42

\bibitem[Dickey \& Lockman(1990)]{Dickey90} Dickey, J.M.\ \& Lockman,
F.J.\ 1990, ARA\&A, 28, 215

\bibitem[Hessels et al.(2006)]{Hessels06} Hessels, J.W.T, Ranson, S.M.,
Stairs, I.H., Freire, C.C., Kaspi, V.M., Camilo, F.\ 2006, Science,
311, 1901

\bibitem[in 't Zand et al.(2001)]{intZand01} in 't Zand, J.J.M.\ et
al.\ 2001, A\&A, 372, 916

\bibitem[Jahoda et al.(2006)]{Jahoda06} Jahoda, K., Markwardt, C.B.,
Radeva, Y., Rots, A.H., Stark, M.J., Swank, J.H., Strohmayer, T.E.,
Zhang, W.\ 2006, ApJS, 163, 401

\bibitem[Kaaret et al.(2002)]{Kaaret02} Kaaret, P., in 't Zand, 
J.J.M.,  Heise, J., Tomsick, J.A.\ 2002, ApJ, 575, 1018

\bibitem[Kaaret et al.(2003)]{Kaaret03} Kaaret, P., in 't Zand, 
J.J.M.,  Heise, J., Tomsick, J.A.\ 2003, ApJ, 598, 481

\bibitem[Kaaret et al.(2006)]{Kaaret06} Kaaret, P., Morgan, E.H.,
Vanderspek, R., Tomsick, J.A.\ 2006, ApJ, 638, 963

\bibitem[Kuulkers et al.(2003)]{Kuulkers03} Kuulkers, E., den Hartog,
P.R., in 't Zand, J.J.M., Verbunt, F.W.M., Harris, W.E., Cocchi, M.\
2003, A\&A, 399, 663

\bibitem[Leahy et al.(1983)]{Leahy83} Leahy, D.A., Darbro, W., Elsner,
R.F., Weisskopf, M.C., Sutherland, P.G., Kahn, S., Grindlay, J.E.\
1983, ApJ, 266, 160

\bibitem[Markwardt et al.(1999)]{Markwardt99} Markwardt, C. B.,
Marshall, F. E., Swank, J. H., \& Wei, C. 1999, IAU Circ. 7300

\bibitem[Revnivtsev, Molkov, Sazonov(2006)]{Revnivtsev06} Revnivtsev,
M., Molkov, S., Sazonov, S.\ 2006, MNRAS, 373, L11

\bibitem[Strohmayer et al.(1996)]{Strohmayer96} Strohmayer, T.E.,
Zhang, W., Swank, J.H., Smale, A., Titarchuk, L., Day, C., Lee, U.
1996, ApJL, 469, L9

\bibitem[Strohmayer \& Markwardt(2002)]{Strohmayer02} Strohmayer, T.E.\
\& Markwardt, C.B.\ 2002, ApJ, 577, 337

\bibitem[Strohmayer et al.(2003)]{Strohmayer03} Strohmayer, T.E.\ \&
Markwardt, C.B., Swank, J.H., in 't Zand, J.\ 2003, ApJ, 596, L67

\bibitem[Torres et al.(2006)]{Torres06} Torres, M. A. P.\ et al.\ 2006,
ATEL 784

\bibitem[van der Klis(1995)]{vdk95} van der Klis, M.\ 1995,
in X-Ray Binaries, ed. W.H.G.\ Lewin, J.\ van Paradijs, \&
E.P.J.\ van den Heuvel (Cambridge: Cambridge Univ. Press),
252

\bibitem[Watts, Strohmayer, Markwardt(2005)]{Watts05} Watts, A.L.,
Strohmayer, T.E., Markwardt, C.B.\ 2005, ApJ, 634, 547

 
\end{thebibliography}
\end{document}